\documentclass[a4,12pt]{article}
\usepackage{graphicx}
\usepackage{amssymb}
\usepackage{slashed}
\usepackage{feynmf}
\usepackage{braket}
\usepackage{empheq}
\usepackage{dcolumn}
\usepackage{color}
\usepackage{comment}
\usepackage[pagebackref=false, colorlinks=true]{hyperref}
\hypersetup{linkcolor=blue, citecolor=blue, urlcolor=blue}
\usepackage{geometry}
\usepackage{authblk}
\usepackage[font=small]{caption}
\usepackage[font=small]{subcaption}
\newcommand{\be}{\begin{eqnarray}}
\newcommand{\ee}{\end{eqnarray}}

\newcommand{\f}{\frac}
\newcommand{\p}{\partial}

\usepackage{tikz}
\usetikzlibrary{arrows.meta,arrows}
\usepackage{soul}

\title{de Sitter corrections to supertranslation Ward identity and soft graviton theorem}
\author[a]{Pratik Chattopadhyay\thanks{pratikpc@gmail.com}}
\author[b]{Divyesh N. Solanki\thanks{divyeshsolanki98@gmail.com}}
\affil[a]{\small School of Physics, The University of Electronic Science and Technology of China, No.2006, Xiyuan Avenue, West Hi-Tech Zone, Chengdu, Sichuan, P.R.China, Post Code: 611731.}
\affil[b]{\small Indian Institute of Information Technology Allahabad (IIITA), Devghat, Jhalwa, Prayagraj-211015, Uttar Pradesh, India.}
\date{}

\begin{document}
\maketitle

\begin{abstract}
We study the tree-level scattering of massless scalars followed by an emission of a soft graviton in the small compact region inside the static patch in de Sitter space. We derive in the small cosmological constant limit the perturbative corrections to the Weinberg soft graviton theorem. Exploiting the remarkable relationship between asymptotic symmetries and soft theorems in flat space, we derive perturbative corrections to the supertranslation Ward identity. We further show that the derived supertranslation Ward identity reduces to the perturbative soft graviton theorem for the same choice of the supertranslation parameter as in flat space.
\end{abstract}

\section{Introduction}
There has been a resurgence of interest in the soft factorization properties of scattering amplitudes, called soft theorems \cite{Bloch_1937, Gell-Mann_Goldberger, Low_1954, Low_1958, Weinberg_1965, Gross_Jackiw, Jackiw_1968, White_2011}, and their connection with the asymptotic symmetries of the underlying gauge theories and gravity, which is given by the Ward identity \cite{Temple_He, Laddha_2015, Daniel_Kapec, He_Lysov_Mitra_Strominger, Lysov_Pasterski_Strominger, Campiglia_Laddha_2016, Kapec_Lysov_2014, Campiglia_Laddha_2014, Campiglia_Laddha_2015}. There is another cornerstone, called memory effect, which has connections with soft theorems and asymptotic symmetries in the deep infrared regime \cite{Strominger_Zhiboedov, Pasterski_Strominger_Zhiboedov, Strominger}. It provides observational signatures of the more abstract theories of asymptotic symmetries and soft theorems. The memory effects have also been studied in the context of near-horizon asymptotic symmetries, an interested reader can see \cite{Donnay_2016, Donnay_PRL_2016, Donnay_2018, Bhattacharjee_2020, Bhattacharjee_2021, Bhat_2025} and references therein. Loop-corrections introduce non-analytic terms in the soft expansion, which depend on the logarithm of the soft energy \cite{Laddha_Sen_2018, SS_2019, SSS_2020}, which have a connection with tail memory effects \cite{Laddha_Sen_2019, SS_2022}. The relation between logarithmic soft theorems and the asymptotic symmetries has been established in \cite{Campiglia_Laddha_2019, Agrawal_Donnay_2024}, and the first principles derivation using the covariant phase space method have been done recently in \cite{Choi_Laddha_Puhm_1, Choi_Laddha_Puhm_2, Choi_Laddha_Puhm_3}.

Since we live in a universe that approximately asymptotes to de Sitter space at late-times where the value of cosmological constant is very small, it becomes interesting to study the infrared properties of gauge theories and gravity in de Sitter space. The soft photon and soft graviton theorems in the small cosmological constant limit using a field theoretic approach have been explored in great detail in \cite{Sayali_Diksha, SCB}. These studies compute de Sitter corrections to the known flat space soft theorems by confining the scattering set up to the small region inside the static patch. By utilizing the remarkable relationship between the soft theorems and asymptotic symmetries, the soft theorems in the full static patch of de Sitter space have also been derived from the Ward identities of the near-horizon symmetries in \cite{Mao_Zhou, Mao_Zhang}. Moreover, the memory effects in de Sitter background have been studied in \cite{Bieri_2016, Chu_2017, Chu_2021, Tolish_Wald_2016}, and the relation between Bondi-Metzner-Sachs (BMS)-like symmetries and memory effects in the static patch of de Sitter space have been explored in \cite{Hamada_2017}. In addition to this, significant progress have been made on the classical soft theorems for a small cosmological constant in the context of anti-de Sitter space \cite{Fernandes_review_2023, Banerjee_2021, Banerjee_Bhattacharjee_2021, Banerjee_2023}.

By considering the tree-level scattering of massive scalars followed by an emission of a soft graviton in the compact region inside the static patch, the perturbative corrections to the flat space soft graviton theorems have been calculated in \cite{SCB} in the small cosmological constant limit. The full soft factor is written as follows:
\begin{align} \label{Eq. 1.1}
A\big(\{p_i&\}, k \big) = \sum\limits_{i=1}^{n} \dfrac{\kappa}{2} \bigg[\dfrac{p_i^\alpha p_i^\beta\varepsilon_{\alpha\beta}}{p_i\cdot k} - i \dfrac{p_i^\alpha\varepsilon_{\alpha\beta}k _\gamma J^{\beta\gamma}}{p_i\cdot k} - \dfrac{m}{l} \dfrac{p_i^\alpha\varepsilon_{\alpha\beta}k_\gamma J^{\beta\gamma}}{(p_i\cdot k)^2} + \dfrac{3m^2}{2l^2} \dfrac{p_i^\alpha p_i^\beta \varepsilon_{\alpha\beta}}{(p_i\cdot k)^3} \nonumber \\ 
&+ \dfrac{1}{2l^2} \dfrac{p_i^\alpha p_i^\beta\varepsilon_{\alpha\beta}}{(p_i\cdot k)^2} + \dfrac{m^2}{l^2} \dfrac{p_i^\alpha p_i^\beta \varepsilon_{\alpha\beta}}{(p_i\cdot k)^3} k\cdot\partial_{p_i} + \dfrac{3}{2l^2} \dfrac{p_i^\alpha p_i^\beta\varepsilon_{\alpha\beta}}{(p_i\cdot k)^2}p_i\cdot \partial_{p_i} + \dfrac{m^2}{2l^2} \dfrac{p_i^\alpha \varepsilon_{\alpha\beta}}{(p_i\cdot k)^2}\partial_{p_i}^\beta \bigg],
\end{align}
where $p_i\ (i=1,...,n)$ are the momenta of the massive scalars, and $k$ is the momentum of the soft graviton. The momentum $k$ in the above expression is defined on the de Sitter background and obeys the following dispersion relation: $k^2=2/l^2$. It was shown in \cite{SCB} by considering the different set of scalar modes that the $\mathcal{O}(l^{-1})$ corrections depend on the choice of modes, and the $\mathcal{O}(l^{-2})$ corrections do not depend on the choice of modes. Thus, $\mathcal{O}(l^{-1})$ corrections are then non-universal, while it was expected that the $\mathcal{O}(l^{-2})$ corrections are universal.

Since the Weinberg soft graviton theorem can be recovered from the supertranslation Ward identity, it is very interesting to analyze whether the perturbative corrections to the supertranslation Ward identity can recover the perturbative corrections to the Weinberg soft graviton theorem. In this paper, we establish the connection between perturbative soft graviton theorem\footnote{By perturbative soft graviton theorem, we refer to the Weinberg soft graviton theorem $+$ perturbative corrections in the limit of large $l$ in de Sitter space.} and the corresponding supertranslation Ward identity. For ease of the calculations, we consider the massless scalar field minimally coupled to the Einstein gravity. Since the massive scalar field mode solution, for which the perturbative soft graviton theorems were derived in \cite{SCB}, is not well-defined in the massless limit, we study the massless scalar field in the de Sitter background in the small cosmological constant limit, and determine the perturbative corrections to the Weinberg soft graviton theorem. To proceed, we first briefly discuss our scattering setup in the following section.

\subsection{Scattering setup}
\label{setup_sec}
\begin{figure}
\centering
\begin{tikzpicture}
\draw[black, very thick] (-2.5,2.5) to node[shift={(0,0.3)}, sloped] {$\mathcal{I}^+$} (2.5,2.5);
\draw[black, very thick] (-2.5,-2.5) to node[shift={(0,-0.3)}, sloped] {$\mathcal{I}^-$} (2.5,-2.5);
\draw[black, very thick] (-2.5,2.5) to node[shift={(-1.1,0)}] {North Pole} (-2.5,-2.5);
\draw[black, very thick] (2.5,2.5) to node[shift={(1.1,0)}] {South Pole} (2.5,-2.5);
\draw[black, very thick] (-2.5,2.5) to node[shift={(-1,1.6)}] {$H^+$} (2.5,-2.5);
\draw[black, very thick] (2.5,2.5) to node[shift={(-1,-1.5)}] {$H^-$} (-2.5,-2.5);
\draw[fill=orange] (-2.5,2.5) -- (0,0) -- (-2.5,-2.5) -- cycle;
\draw[black, very thick,fill=lightgray] (-2.5,1.5) to [out=-45, in=45,looseness=2] node[shift={(-0.8,0)}] {$R$} (-2.5,-1.5);
\end{tikzpicture}
\caption{\footnotesize The Penrose diagram of the de Sitter space. The scattering processes are confined to the compact region depicted in gray inside the static patch.}
\label{dS_Penrose}
\end{figure}
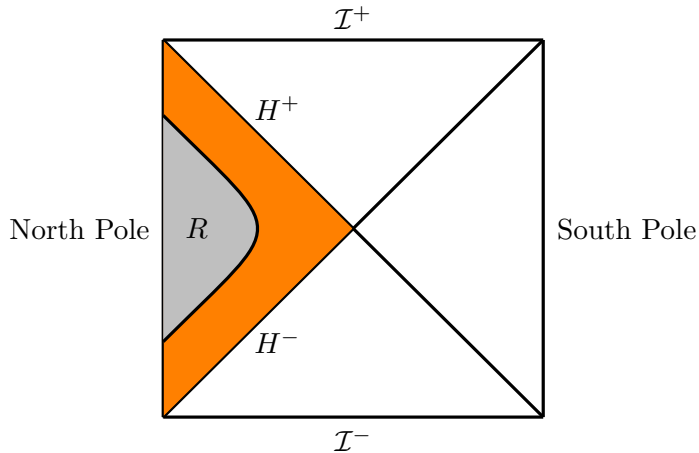

We study the scattering of massless scalars with an emission of a soft graviton in the static patch of the de Sitter space. We confine this scattering process to the small compact region $R$ inside the static patch as shown in Fig. (\ref{dS_Penrose}). The de Sitter metric can be put in the conformally flat form in the stereographic coordinates $x^\mu$ as \cite{SCB, Aldrovandi_1995}
\be 
g_{\mu\nu}=\Omega^2\eta_{\mu\nu}, ~~\Omega=\frac{1}{1+x^2/4l^2},
\ee
where $\eta_{\mu\nu}$ is the Minkowski metric. Inside the compact region, $x^\mu << l$, we can approximate the de Sitter metric up to $\mathcal{O}(l^{-2})$ as follows:
\be 
\label{metric}
g_{\mu\nu}\approx\Big(1-\frac{x^2}{2l^2}\Big)\eta_{\mu\nu}.
\ee
Thus, we effectively work in the large curvature length or small cosmological constant limit. We now discuss the key assumptions on our setup and various scales involved in the scattering process.
\begin{itemize}
\item The region $R$ inside the static patch is much smaller than the de Sitter curvature length $l$ so that the gravitational effects can be treated perturbatively. Also, the region $R$ is much larger than the Planck length so that the gravitational interaction becomes weak as particles reach the boundary. Thus, the scattering takes place in the regime of $\sqrt{G}<<R<<l$, where $G$ is Newton's constant.
\item Since the region $R$ is finite, particles can reach the boundary in a finite time $T$, which should be much larger than the interaction timescale so that the particles are free from the interactions at early and late times. Thus, one can define free particle states on early- and late-time Cauchy slices, which are the Hilbert spaces of the incoming and outgoing states respectively.
\item The $\mathcal{S}$-matrix for the scattering of massless scalars of momenta $p_1,...,p_n$ with an emission of a graviton of momentum $k\ (=\omega\hat{k})$ decomposes in the large curvature length limit and in the soft limit $\omega\to 0$ as follows \cite{SCB}:
\begin{equation} \label{Gamma_structure}
    \Gamma_{n+1}(\{p_1,...,p_n\},\omega\hat{k}) = A(\{p_i\},\omega\hat{k})\  \Gamma_{n}(\{p_1,...,p_n\}),
\end{equation}
\begin{equation}
    A(\{p_i\},\omega\hat{k}) = A^{\text{(flat)}}(\{p_i\},\omega\hat{k}) + \dfrac{1}{l^2} A^{\text{(dS)}}(\{p_i\},\omega\hat{k}),
\end{equation}
where $\Gamma_n$ is the $\mathcal{S}$-matrix for the scattering of massless $n$-scalars without an emission of the graviton, $A^{\text{(flat)}}$ is the flat space soft factor, and $A^{\text{(dS)}}$ contains de Sitter corrections to the flat space soft factor. These corrections are perturbatively small only for $\omega l >>1$. Also, the soft limit in the flat space requires $\omega << E$, where $E$ is the energy of the hard particles that participate in the scattering process. Thus, in this setup, the soft limit is valid in the regime of $\frac{1}{l}<<\omega << E$. Since $l$ is very large and $\omega\to 0$, it is convenient to introduce a constant parameter\footnote{We treat this large constant parameter independent of $\omega$.} $\delta=\omega l$. In this new parameter, the soft factor can be expanded as follows:
\begin{equation}
    A(\{p_i\},\omega\hat{k}) = A^{\text{(flat)}}(\{p_i\},\omega\hat{k}) + \dfrac{1}{\delta^2} \tilde{A}^{\text{(dS)}}(\{p_i\},\omega\hat{k}),
\end{equation}
where $\tilde{A}^{\text{(dS)}} = \omega^2 A^{\text{(dS)}}$.
\end{itemize}

The paper is organized as follows: In section (\ref{massless_scalars_sec}), we study the massless scalar field in the de Sitter background. Particularly, we find an orthogonal set of mode solutions to the massless scalar field equation in the limit of large curvature length in section (\ref{scalar_modes_sec}). We use the orthogonal set of modes to define the scalar field propagator and verify that it is indeed the Green's function of the massless scalar field equation in section (\ref{scalar_propagator_sec}). In section (\ref{graviton_sec}), we briefly review the gravitons in de Sitter background, particularly the mode solution, that have been studied in detail in \cite{SCB}. We then consider the massless scalar field minimally coupled to the Einstein gravity in the de Sitter background, and define the $\mathcal{S}$-matrix for the scattering process depicted in Fig. (\ref{soft_graviton_fig}) in section (\ref{scattering_sec}). Evaluating the $\mathcal{S}$-matrix, we determine perturbative corrections to the Weinberg soft graviton theorem in section (\ref{perturbative_correction_sec}) and summarize results in section (\ref{results_sec}). In section (\ref{symmetry_sec}), we first represent the perturbative soft graviton theorem in the holomorphic coordinates, and briefly review the flat space supertranslation Ward identity in section (\ref{Ward_identity_review}). In section (\ref{perturbative_Ward_sec}), we derive perturbative corrections to the supertranslation Ward identity by exploiting the relationship between asymptotic symmetries and soft theorems. We then show in section (\ref{Ward_to_soft_sec}) that the perturbative soft graviton theorem can be recovered from the supertranslation Ward identity derived in this paper. We finally conclude our work in section (\ref{discussion_sec}).

\section{Massless Scalars in de Sitter Background}
\label{massless_scalars_sec}
In this section, we determine the orthogonal set of mode solutions for the massless scalar field and obtain the scalar field propagator.
\subsection{Scalar modes}
\label{scalar_modes_sec}
The mode solutions for the massive scalar field in de Sitter background in the small cosmological constant limit have already been derived in \cite{Sayali_Diksha} and briefly reviewed in \cite{SCB}, whose massless limit is not well-defined. Thus, in this section, we consider the massless scalar field on the de Sitter background and derive the mode solution. The scalar field equation of motion on the de Sitter background is 
\begin{equation}
    \nabla^2\phi = 0,
\end{equation}
where $\nabla^2$ is the d'Alembertian operator in the de Sitter space. The above scalar field equation in the stereographic coordinates in the large curvature length limit takes the form as
\be \label{EOM}
\left(1+\frac{x^2}{2l^2}\right)\Box\phi -\frac{1}{l^2}x\cdot\partial\phi=0,
\ee
where $\Box$ is the flat space d'Alembertian operator. The scalar field mode expansion can be done as follows:
\be
\phi(x)=\int \f{d^{3}p}{(2\pi)^{3}} \ [a_p\ g_p(x)\ +\ a^\dagger_p\ g^*_p (x)],
\ee
where $a_p\ (a^\dagger_p)$ annihilates (creates) a scalar particle with the momentum $p$, and $g_p(x)$ are the scalar field modes parametrized by the momentum $p$. To determine the mode solution, we choose the following ansatz:
\be \label{ansatz}
g_p(x) = \dfrac{e^{ip\cdot x}}{\sqrt{2E_p}} \left(1 + \dfrac{\mathcal{F}(x)}{l^2} \right),
\ee
where $E_p$ is the energy of the scalar field, and $\mathcal{F}(x)$ is an arbitrary function which we want to determine. Substituting (\ref{ansatz}) into (\ref{EOM}) as a solution of the scalar field $\phi$, we obtain:
$$\mathcal{F}(x)=\dfrac{x^2}{4},\ \ p^2=\dfrac{2}{l^2}.$$
Now, since we have obtained the mode solutions, we need to verify whether they are orthogonal or not. We choose two modes, one parametrized by the momentum $p$ and another parametrized by the momentum $q$, and define their inner product on the solution space as \cite{Sayali_Diksha, SCB}:
\begin{equation}
\label{scalar_inner_product}
    \left( g_p, g_q \right) = -i \int d^3x \sqrt{-g} \ \big(g^*_p(x) \nabla^t g_q(x) - g_q(x) \nabla^t g^*_p(x) \big).
\end{equation}
Substituting the mode solutions (\ref{ansatz}) in the above expression, one can verify that
\begin{equation}
    \left( g_p, g_q \right) = (2\pi)^3 \delta^{(3)}(p-q).
\end{equation} 
Thus, the scalar field mode solutions obtained in this section are indeed orthogonal.

\subsection{Scalar propagator}
\label{scalar_propagator_sec}
In this section, we determine the scalar field propagator on the de Sitter background in the large curvature length limit. Since the propagators are the Green's functions of the equations of motion, the differential equation for the scalar field propagator can be written as
\be
\nabla_x^2 D(x,y)=i \frac{\delta^{(4)}(x-y)}{\sqrt{-g}},
\ee
which can be expanded in the stereographic coordinates up to $\mathcal{O}(l^{-2})$ as
\be \label{propeom}
\left[\left(1+\frac{x^2}{2l^2}\right)\Box_x-\frac{1}{l^2}x\cdot\p_x\right]D(x,y) = i \frac{\delta^{(4)}(x-y)}{\sqrt{-g}}.
\ee
We can write the symmetric form of the scalar field propagator using the orthogonal set of modes $g_p$ as follows:
\begin{align} \label{scalar_propagator}
    D(x,y)&=-i\int\dfrac{d^4p}{(2\pi)^4} \dfrac{(\sqrt{2E_p})g_p(x)(\sqrt{2E_p})g^*_p(y)}{p^2-\frac{2}{l^2}-i\epsilon} \nonumber \\
    &=-i\int\dfrac{d^4p}{(2\pi)^4} \dfrac{e^{ip\cdot(x-y)}}{p^2} \left(1 + \dfrac{x^2}{4l^2} + \dfrac{y^2}{4l^2} + \dfrac{2}{p^2l^2} \right).
\end{align}
Now, we need to verify whether the above form of the propagator satisfies the differential equation (\ref{propeom}). Let us substitute Eq. (\ref{scalar_propagator}) into (\ref{propeom}):
\begin{align}
\left[\left(1+\frac{x^2}{2l^2}\right)\Box_x-\frac{1}{l^2}x\cdot\p_x\right]D(x,y) &= i \int \dfrac{d^4p}{(2\pi)^4} \left(1+\dfrac{3x^2}{4l^2}+\dfrac{y^2}{4l^2} \right) e^{ip\cdot(x-y)} \nonumber \\
&= i \left(1+\dfrac{3x^2}{4l^2}+\dfrac{y^2}{4l^2} \right) \delta^{(4)}(x-y) \nonumber \\
&= i \left(1+\dfrac{y^2}{l^2} \right) \delta^{(4)}(x-y) \nonumber \\
&= i \dfrac{\delta^{(4)}(x-y)}{\sqrt{-g(y)}}.
\end{align}
Hence, the scalar field propagator (\ref{scalar_propagator}) obtained from the orthogonal set of scalar modes is indeed the Green's function of the massless scalar field equation of motion.

\section{Gravitons in de Sitter background: review}
\label{graviton_sec}
In this section, we briefly review gravitons in the de Sitter background, which have already been studied extensively in \cite{SCB}. We particularly define the mode solution as it is essential in the computation of the $\mathcal{S}$-matrix.
\subsection{Mode expansion}
The linearized Einstein field equation in the de Sitter background in transverse-traceless (TT) gauge, $\nabla_\mu h^{\mu\nu}=0=h$, is defined as 
\begin{equation}
\label{linearized_Einstein_dS}
\left(\nabla^2 - \dfrac{2}{l^2} \right) h_{\mu\nu} = 0.
\end{equation}
The above linearized field equation is invariant under the following residual gauge transformations \cite{Date_Hoque_2016}:
\be
h_{\mu\nu} \to h_{\mu\nu} + \nabla_{\mu}\xi_{\nu} + \nabla_{\nu}\xi_{\mu},
\ee
where the vector fields $\xi_\mu$ must satisfy
\begin{equation}
    \left(\nabla^2 + \dfrac{3}{l^2} \right) \xi_\mu = 0, \ \ \ \ \nabla_\mu \xi^\mu = 0.
\end{equation}
The mode expansion of the field $h_{\mu\nu}$ can be defined as \cite{SCB}
\begin{equation}
    h_{\mu\nu}(x) = \sum_{h=\pm} \int\dfrac{d^3 k}{(2\pi)^3} \left(a^h_k f^h_{\mu\nu}(x,k) + a^{h\dagger}_k f^{h*}_{\mu\nu}(x,k) \right),
\end{equation}
where $a^h_k$ ($a^{h\dagger}_k$) is the annihilation (creation) operator which acts on the asymptotic states to annihilate (create) a graviton of momentum $k$ and helicity ``$h$."\footnote{The helicity index ``$h$" should not be confused with the trace of $h_{\mu\nu}$.}
The graviton mode solution $f^h_{\mu\nu}(x,k)$ has already been obtained in \cite{SCB} by considering a suitable ansatz and using it to solve the linearized field equation. The following is the solution
\begin{align}
\label{graviton_mode}
    &f^h_{\mu\nu}(x,k) = \dfrac{e^{ik\cdot x}}{\sqrt{2E_k}} \left(\varepsilon^h_{\mu\nu} + \dfrac{1}{2l^2} \varepsilon^h_{\mu\alpha}x^\alpha x_\nu + \dfrac{1}{2l^2} \varepsilon^h_{\alpha\nu}x^\alpha x_\mu - \dfrac{1}{4l^2} \varepsilon^h_{\mu\nu}x^2 - \dfrac{1}{4l^2} \varepsilon^h_{\alpha\beta}x^\alpha x^\beta \eta_{\mu\nu} \right), \nonumber \\
    &k^2 = \dfrac{2}{l^2};
\end{align}
where $\varepsilon_{\mu\nu}$ is the polarization tensor, and $E_k$ is the zeroth component of $k^\mu$. The mode solution also satisfies the TT gauge condition and yields $k^\mu\varepsilon_{\mu\nu}=\mathcal{O}(l^{-2}),\ \varepsilon^{\mu}_{\mu}=0$. To check whether the modes are orthogonal, we define the inner product on the solution space as \cite{SCB}
\begin{multline}
\label{graviton_inner_product}
    \left(\mathcal{P}^{\mu\nu\alpha\beta} f^h_{\mu\nu}(k), f^{h'}_{\alpha\beta}(k') \right) = -i \int d^3x \sqrt{-g} \  \mathcal{P}^{\mu\nu\alpha\beta}\ e^{-\epsilon\frac{|\overrightarrow{x}|}{R}} \big(f^{*h}_{\mu\nu}(x,k) \nabla^t f^{h'}_{\alpha\beta}(x,k') \\- f^{h'}_{\mu\nu}(x,k') \nabla^t f^{*h}_{\alpha\beta}(x,k) \big),
\end{multline}
\begin{equation}
    \mathcal{P}^{\mu\nu\alpha\beta} = \dfrac{1}{2}\left(g^{\mu\alpha} g^{\nu\beta} + g^{\mu\beta} g^{\nu\alpha} - g^{\mu\nu}g^{\alpha\beta} \right);
\end{equation}
where the exponential damping factor was introduced to get rid of the boundary terms. Plugging the modes (\ref{graviton_mode}) parametrized by two different momenta $k$ and $k'$ in the inner product expression above, we obtained
\begin{equation}
    \left(\mathcal{P}^{\mu\nu\alpha\beta} f^h_{\mu\nu}(k), f^{h'}_{\alpha\beta}(k') \right) = (2\pi)^3 \delta_{hh'} \delta^{(3)}(k-k').
\end{equation}
Thus, the graviton modes are the orthogonal set of modes.

\section{Soft graviton scattering}
\label{scattering_sec}
In this section, we study the scattering of massless scalars with an emission of a soft graviton. We define the corresponding $\mathcal{S}$-matrix and obtain the perturbative corrections to the soft graviton theorem.

The action for a massless scalar field minimally coupled to the Einstein gravity is given by
\begin{equation}
S = -\int d^4x \sqrt{-g} \bigg[\frac{2}{\kappa^2}(\mathcal{R} - 2\Lambda)+ \frac{1}{2}g^{\mu\nu} \nabla_\mu\phi \nabla_\nu\phi + V(\phi) \bigg],
\end{equation}
where $\mathcal{R}$ is the Ricci scalar, $\Lambda\ (=3/l^2)$ is a cosmological constant, $V(\phi)$ is a scalar potential, $g_{\mu\nu}$ is a general spacetime metric, and $\nabla_\mu$ is the covariant derivative compatible with $g_{\mu\nu}$. In the weak field expansion, $g_{\mu\nu} = \Bar{g}_{\mu\nu} + \kappa h_{\mu\nu}$,\footnote{Bar notations are used for the de Sitter space to distinguish it from the general spacetime, which will be dropped later.} the leading order terms are
\begin{align}
    \mathcal{L}_{\text{gravity}} =& -\frac{2}{\kappa^2}\sqrt{-g} (\mathcal{R} - 2\Lambda) \nonumber \\
    =&\ \frac{1}{2}\Bar{\nabla}_\sigma h_{\mu\nu} \Bar{\nabla}^\sigma h^{\mu\nu} - \frac{1}{2}\Bar{\nabla}_\sigma h \Bar{\nabla}^\sigma h - \Bar{\nabla}_\mu h_{\nu\sigma} \Bar{\nabla}^\sigma h^{\mu\nu} + \Bar{\nabla}_\mu h^{\mu\nu} \Bar{\nabla}_\nu h \nonumber\\ 
    &- \frac{3}{l^2} \bigg(h_{\mu\nu}h^{\mu\nu} - \frac{1}{2}h^2 \bigg) + \cdots ,
\end{align}
\begin{align}
    \mathcal{L}_{\text{scalar}} &= -\sqrt{-g} \bigg[\frac{1}{2}g^{\mu\nu} \nabla_\mu\phi \nabla_\nu\phi+ V(\phi) \bigg] \nonumber \\
    &= -\frac{1}{2} \Bar{\nabla}_\sigma\phi \Bar{\nabla}^\sigma\phi- V(\phi) + \frac{\kappa}{2} h^{\mu\nu} \bigg[\Bar{\nabla}_\mu\phi \Bar{\nabla}_\nu\phi - \frac{1}{2} \Bar{g}_{\mu\nu} \big(\Bar{\nabla}_\sigma\phi \Bar{\nabla}^\sigma\phi\big)  \bigg] + \cdots ,
\end{align}
where the total derivative terms and the constant terms have been dropped in the Lagrangian.

Let us now consider the scattering of massless $n$-scalars with momenta $p_i\ (i=1,...,n)$ followed by an emission of a graviton with momentum $k$. There are two possible ways for the emission of the graviton, either from an external line or from an internal line as shown in Fig. (\ref{soft_graviton_fig}). The total $\mathcal{S}$-matrix for such a scattering process has already been defined in \cite{SCB}, which we simply write here as follows:
\begin{figure}
    \centering
    \includegraphics[width=0.35\linewidth]{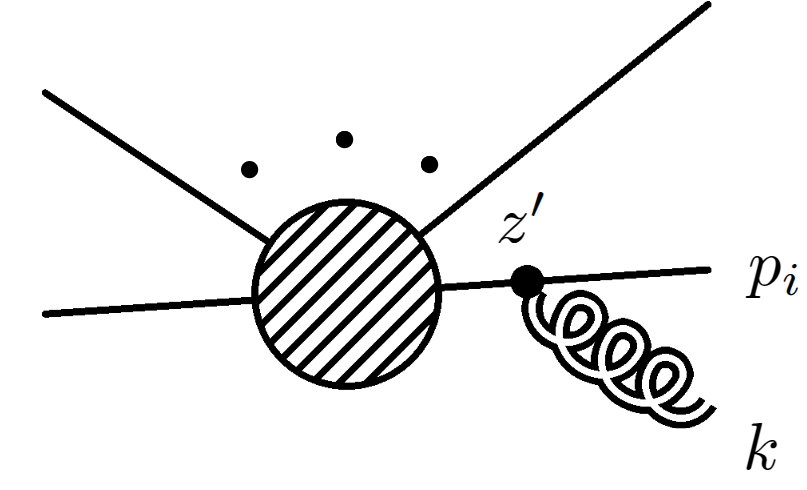}
    \hspace{20pt}
    \includegraphics[width=0.34\linewidth]{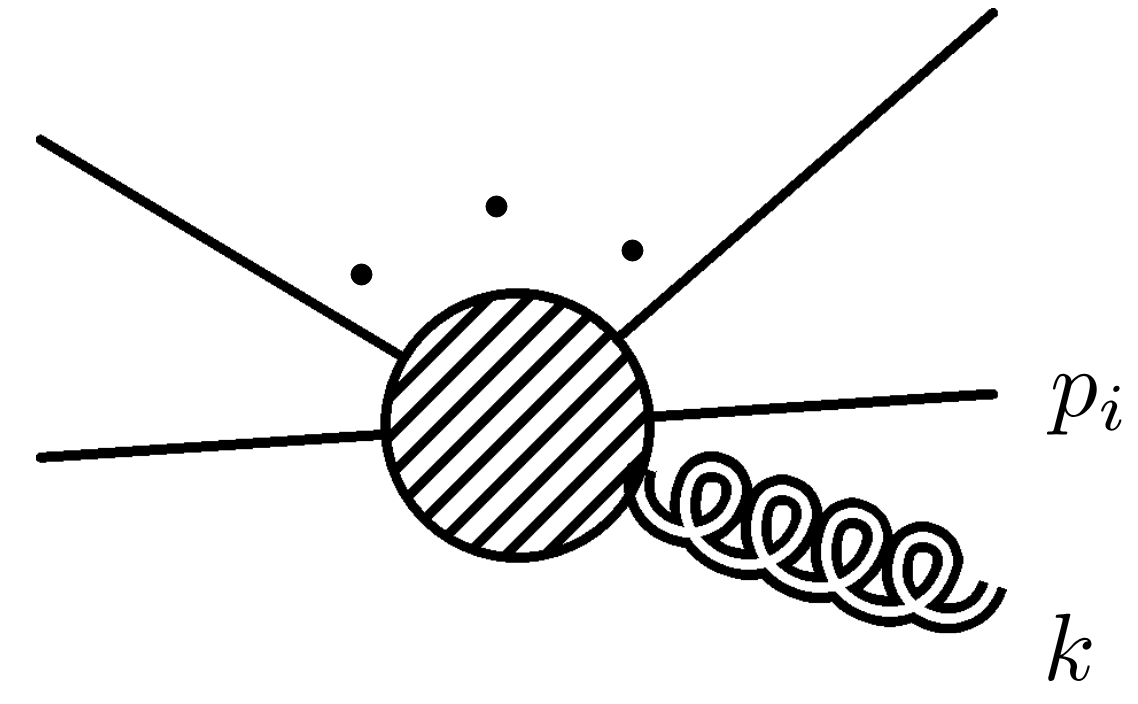}
    \caption{\footnotesize Scattering of massless scalars with momenta $p_i$, emitting a graviton with momentum $k$.}
    \label{soft_graviton_fig}
\end{figure}
\begin{align} \label{Gamma_LSZ}
&\Gamma_{n+1}(\{p_1,...,p_n\},k) = \sum_{i=1}^{n}\frac{\kappa}{2} \int d^4z_i \sqrt{-g(z_i)} g^{*}_{p_i}(z_i) (-i) (\nabla^{2}_{z_i}) \nonumber \\
&\cdot \int d^4y \sqrt{-g(y)} \mathcal{P}^{\mu\nu\alpha\beta} f^{*}_{\mu\nu}(y,k) (-i) (\nabla_y^2 - 2/l^2) \int\prod_{j=1; j\ne i}^{n-1} d^4z_j \sqrt{-g(z_j)} g^*_{p_j}(z_j)(-i)(\nabla^{2}_{z_j}) \nonumber \\ 
&\cdot \int d^4z \sqrt{-g(z)} \int d^4z' \sqrt{-g(z')}\ \bigg\langle \mathcal{T} \phi(z_i) h_{\alpha\beta}(y) h^{\rho\sigma}(z') \nonumber\\ &\cdot\bigg(\nabla'_{\rho}\phi(z') \nabla'_{\sigma}\phi(z') - \frac{1}{2}g_{\rho\sigma} \nabla'_{\tau}\phi(z') \nabla'^{\tau}\phi(z')\bigg) G[\phi(z)...\phi(z_j)] \bigg\rangle + N_{n+1}(\{p_1,...,p_n\},k),
\end{align}
where $G[\phi(z)...\phi(z_j)]$ is the $n$-point correlation function which receives contributions from any arbitrary interaction vertices, $N_{n+1}$ is the $\mathcal{S}$-matrix for the second diagram in Fig. (\ref{soft_graviton_fig}) where the graviton is not attached to an external line, and $\kappa$ is the scalar-graviton coupling constant. The summation has been taken over all the scalars since any of them can emit the graviton. The above expression can be simplified in the same way as we did in \cite{SCB}: We first perform the Wick contraction among the fields, which yields respective propagators, i.e. scalar field and graviton propagators. The equation of motion operators, $\nabla^2_{z_i}$ and $(\nabla^2_y-2/l^2)$ in (\ref{Gamma_LSZ}), now act on the respective propagators and yield the delta functions. Then the $z_i$- and $y$-integrals can be performed trivially on the delta functions. The resultant expression can be further simplified by using the traceless condition of the graviton modes, i.e. $g^{\mu\nu}f^*_{\mu\nu}=0$. Finally, one obtains the following expression:
\begin{align} \label{LSZ_final}
&\Gamma_{n+1}(\{p_1,...,p_n\},k) = \sum_{i=1}^{n}\frac{\kappa}{2} \int d^4z d^4z' \left(1-\dfrac{z^2}{l^2} - \dfrac{z'^2}{l^2} \right) f^{*\rho\sigma}(z',k) \bigg\{2\partial'_\rho g^*_{p_i}(z') \partial'_\sigma D(z',z) \bigg\} \nonumber \\
&\cdot\int\prod_{j=1; j\ne i}^{n-1} d^4z_j \sqrt{-g(z_j)} g^*_{p_j}(z_j)(-i)(\nabla^{2}_{z_j}) \big\langle G[\{\phi(z_j) \}] \big\rangle + N_{n+1}(\{p_1,...,p_n\},k).
\end{align}
This is the main expression from which we will derive the perturbative corrections to the Weinberg soft graviton theorem in the following section.

\subsection{Perturbative corrections to Weinberg soft graviton theorem}
\label{perturbative_correction_sec}
Let us now substitute scalar modes (\ref{ansatz}), scalar propagator (\ref{scalar_propagator}), and graviton modes (\ref{graviton_mode}) into Eq. (\ref{LSZ_final}). We obtain
\begin{align}
&\Gamma_{n+1}(\{p_1,...,p_n\},k) = -i\sum_{i=1}^{n}\kappa
\int d^4z d^4z' \int\dfrac{d^4p}{(2\pi)^4} \bigg(\dfrac{p^\alpha p_i^\beta \varepsilon_{\alpha\beta}}{p^2} + \dfrac{z'^2(p^\alpha p_i^\beta \varepsilon_{\alpha\beta})}{4l^2p^2} \nonumber \\
&- \dfrac{3z^2(p^\alpha p_i^\beta \varepsilon_{\alpha\beta})}{4l^2p^2}
+ \dfrac{i(p^\alpha z'^\beta \varepsilon_{\alpha\beta})}{2l^2p^2} + \dfrac{(p_i\cdot z')(p^\alpha z'^\beta \varepsilon_{\alpha\beta})}{2l^2p^2} - \dfrac{i(p_i^\alpha z'^\beta \varepsilon_{\alpha\beta})}{2l^2p^2} + \dfrac{(p\cdot z')(p_i^\alpha z'^\beta \varepsilon_{\alpha\beta})}{2l^2p^2} \nonumber \\
&- \dfrac{(p\cdot p_i)(z'^\alpha z'^\beta \varepsilon_{\alpha\beta})}{4l^2p^2} + \dfrac{2(p^\alpha p_i^\beta \varepsilon_{\alpha\beta})}{l^2p^4} \bigg) e^{i(p-p_i-k)\cdot z'} e^{-ip\cdot z} \nonumber \\
&\cdot\int\prod_{j=1; j\ne i}^{n-1} d^4z_j \sqrt{-g(z_j)} g^*_{p_j}(z_j)(-i)(\nabla^{2}_{z_j}) \big\langle G[\{\phi(z_j) \}] \big\rangle + N_{n+1}(\{p_1,...,p_n\},k).
\end{align}
The above expression can be simplified in the following way: we first replace all $z'$ by $-i\partial_p$ and perform the integration by parts in $p$, which brings the exponential factor $e^{ip\cdot z'}$ outside the derivative operator $\partial_p$. We then collect all the exponential factors to $e^{i(p-p_i-k)\cdot z'}$ and do the $z'$-integral which yields the delta function $\delta^{(4)}(p-p_i-k)$. We finally perform the $p$-integral on the delta function and take the soft limit $k\to 0$.
\begin{align}
&\Gamma_{n+1}(\{p_1,...,p_n\},k) = -i\sum_{i=1}^{n} \dfrac{\kappa}{2} 
\int d^4z \bigg(\dfrac{p_i^\alpha p_i^\beta \varepsilon_{\alpha\beta}}{p_i\cdot k} - \dfrac{i(k\cdot z)(p_i^\alpha p_i^\beta \varepsilon_{\alpha\beta})}{p_i\cdot k} + \dfrac{1}{2l^2} \dfrac{p_i^\alpha p_i^\beta \varepsilon_{\alpha\beta}}{(p_i\cdot k)^2} \nonumber \\
&-\dfrac{3i}{2l^2} \dfrac{(p_i\cdot z)(p_i^\alpha p_i^\beta \varepsilon_{\alpha\beta})}{(p_i\cdot k)^2}
\bigg) e^{-ip_i\cdot z} \int\prod_{j=1; j\ne i}^{n-1} d^4z_j \sqrt{-g(z_j)} g^*_{p_j}(z_j)(-i)(\nabla^{2}_{z_j}) \big\langle G[\{\phi(z_j) \}] \big\rangle \nonumber \\
& + N_{n+1}(\{p_1,...,p_n\},k).
\end{align}
We now replace $z$ by $i\partial_{p_i}$, and put the above expression in the final form as
\begin{align} \label{Gamma=soft+N}
&\Gamma_{n+1}(\{p_1,...,p_n\},k) = \sum_{i=1}^{n} \dfrac{\kappa}{2} \bigg(\dfrac{p_i^\alpha p_i^\beta \varepsilon_{\alpha\beta}}{p_i\cdot k} + \dfrac{p_i^\alpha p_i^\beta \varepsilon_{\alpha\beta}}{p_i\cdot k} k\cdot\partial_{p_i} + \dfrac{1}{2l^2} \dfrac{p_i^\alpha p_i^\beta \varepsilon_{\alpha\beta}}{(p_i\cdot k)^2}\nonumber \\ 
&+ \dfrac{3}{2l^2} \dfrac{p_i^\alpha p_i^\beta \varepsilon_{\alpha\beta}}{(p_i\cdot k)^2} p_i\cdot\partial_{p_i}
\bigg) \Gamma_{n}(p_1,...,p_n) + N_{n+1}(\{p_1,...,p_n\},k),
\end{align}
where $\Gamma_n$ is the $\mathcal{S}$-matrix for the scattering of massless $n$-scalars without an emission of the graviton. The above expression should be invariant under the gauge transformation,
\begin{equation}
    \varepsilon_{\mu\nu} \to \varepsilon_{\mu\nu} + i k_\mu \xi_\nu + i k_\nu \xi_\mu + \mathcal{O}(l^{-2}).
\end{equation}
Using the gauge-invariance property of soft factors (or total $\mathcal{S}$-matrix), one can determine the form of $N_{n+1}$, which is decomposed as \cite{SCB}
\begin{equation} \label{N_decomposition}
    N_{n+1}(\{p_1,...,p_n\},k) = -\dfrac{1}{2} \sum_{i=1}^{n} \dfrac{\kappa}{2} \varepsilon_{\alpha\beta} \big(p_i^\alpha \partial^\beta_{p_i} + p_i^\beta \partial^\alpha_{p_i} \big) \Gamma_{n}(p_1,...,p_n) + \dfrac{1}{l^2} N^{\text{(dS)}}_{n+1}(\{p_1,...,p_n\},k).
\end{equation}
Substituting Eq. (\ref{N_decomposition}) into (\ref{Gamma=soft+N}) and simplifying it, we obtain the final expression as follows:
\begin{align}
&\Gamma_{n+1}(\{p_1,...,p_n\},k) = \sum_{i=1}^{n} \dfrac{\kappa}{2} \bigg(\dfrac{p_i^\alpha p_i^\beta \varepsilon_{\alpha\beta}}{p_i\cdot k} - i\dfrac{p_i^\alpha \varepsilon_{\alpha\beta}k_\gamma J_i^{\beta\gamma}}{p_i\cdot k} + \dfrac{1}{2l^2} \dfrac{p_i^\alpha p_i^\beta \varepsilon_{\alpha\beta}}{(p_i\cdot k)^2} \nonumber \\
&+ \dfrac{3}{2l^2} \dfrac{p_i^\alpha p_i^\beta \varepsilon_{\alpha\beta}}{(p_i\cdot k)^2} p_i\cdot\partial_{p_i} \bigg) \Gamma_{n}(p_1,...,p_n) + \dfrac{1}{l^2} N^{\text{(dS)}}_{n+1}(\{p_1,...,p_n\},k),
\end{align}
where $J^{\alpha\beta}=i\big(p^\alpha\partial^\beta_p - p^\beta\partial^\alpha_p \big)$ is the angular momentum operator. We note that the above expression is consistent with the massless limit of (\ref{Eq. 1.1}), even though it was derived from a completely different set of scalar field modes. It gives a strong hint towards the universality of the perturbative corrections to the flat space soft theorem in Eq. (\ref{Eq. 1.1}).

Since the de Sitter curvature length is extremely large, it is now convenient to introduce a constant parameter $\delta=\omega l$ in the soft limit $\omega\to 0$. Therefore,
\begin{align}
\label{Gamma_in_delta_1}
&\Gamma_{n+1}(\{p_1,...,p_n\},\omega \hat{k}) = \sum_{i=1}^{n} \dfrac{\kappa}{2} \bigg(\dfrac{p_i^\alpha p_i^\beta \varepsilon_{\alpha\beta}}{\omega p_i\cdot \hat{k}} - i\dfrac{p_i^\alpha \varepsilon_{\alpha\beta} \hat{k}_\gamma J_i^{\beta\gamma}}{p_i\cdot \hat{k}} + \dfrac{1}{2\delta^2} \dfrac{p_i^\alpha p_i^\beta \varepsilon_{\alpha\beta}}{(p_i\cdot \hat{k})^2} \nonumber \\ 
&+ \dfrac{3}{2\delta^2} \dfrac{p_i^\alpha p_i^\beta \varepsilon_{\alpha\beta}}{(p_i\cdot \hat{k})^2} p_i\cdot\partial_{p_i}
\bigg) \Gamma_{n}(p_1,...,p_n) + \dfrac{1}{\delta^2} N^{\text{(dS)}}_{n+1}(\{p_1,...,p_n\},\hat{k}).
\end{align}
We must note in the above expression that the momenta $p_i$ and $k$ are defined on the de Sitter background and obey the dispersion relation $p_i^2=k^2=2/l^2$. Thus, there are some hidden correction terms that emerge when we decompose the de Sitter momenta into the flat space momenta $p'_i$ and $k'$ that obey $p'^2_i=k'^2=0$. Using the dispersion relations, we can parametrize the de Sitter momenta as follows:
\begin{align}
    &k^\mu =  \omega \left(1,\ \left(1+\dfrac{1}{\delta^2} \right) {\mathbf{\hat{k}}} \right), \\
    &p_i^\mu = E_i \left(1,\ \left(1+\dfrac{\omega^2}{\delta^2 E_i^2} \right) {\mathbf{\hat{p_i}}} \right),
\end{align}
where $\mathbf{\hat{k}}$ and $\mathbf{\hat{p_i}}$ are the unit 3-vectors that parametrize points on the sphere. Now, the de Sitter momenta can be expressed in the flat space momenta as
\begin{align}
    &(k^\mu)_{\text{dS}} \to (k^\mu)_{\text{flat}} + \dfrac{1}{\delta^2} (\vec{k})_{\text{flat}}, \\
    &(p_i^\mu)_{\text{dS}} \to (p^\mu_i)_{\text{flat}} + \dfrac{\omega^2}{\delta^2 E^2_i} (\vec{p_i})_{\text{flat}}.
\end{align}
Thus, Eq. (\ref{Gamma_in_delta_1}) takes the final form in flat space momenta as
\begin{align}
\label{Gamma_in_delta_final}
&\Gamma_{n+1}(\{p_1,...,p_n\},\omega \hat{k}) = \sum_{i=1}^{n} \dfrac{\kappa}{2} \bigg(\dfrac{p_i^\alpha p_i^\beta \varepsilon_{\alpha\beta}}{\omega p_i\cdot \hat{k}} - i\dfrac{p_i^\alpha \varepsilon_{\alpha\beta} \hat{k}_\gamma J_i^{\beta\gamma}}{p_i\cdot \hat{k}} + \dfrac{1}{2\delta^2} \dfrac{p_i^\alpha p_i^\beta \varepsilon_{\alpha\beta}}{(p_i\cdot \hat{k})^2} + \dfrac{3}{2\delta^2} \dfrac{p_i^\alpha p_i^\beta \varepsilon_{\alpha\beta}}{(p_i\cdot \hat{k})^2} p_i\cdot\partial_{p_i} \nonumber \\
&- \dfrac{1}{\delta^2} \dfrac{(p_i^\alpha p_i^\beta \varepsilon_{\alpha\beta})(\Vec{p}_i\cdot \mathbf{\hat{k}})}{\omega (p_i\cdot \hat{k})^2} + \dfrac{1}{\delta^2} \dfrac{p_i^\alpha p_i^\beta \varepsilon_{\alpha\beta}}{p_i\cdot\hat{k}} \mathbf{\hat{k}}\cdot\partial_{\Vec{p}_i} - \dfrac{1}{\delta^2} \dfrac{(p_i^\alpha p_i^\beta \varepsilon_{\alpha\beta})(\vec{p}_i\cdot\mathbf{\hat{k}})}{(p_i\cdot\hat{k})^2} \hat{k}\cdot\partial_{p_i}
\bigg) \Gamma_{n}(p_1,...,p_n) \nonumber \\ 
&+ \dfrac{1}{\delta^2} N^{\text{(dS)}}_{n+1}(\{p_1,...,p_n\}, \hat{k}),
\end{align}
where the prime-notations for the flat space momenta have been dropped. Hereafter, all the momenta are the flat space momenta satisfying $p_i^2=0=k^2$. 

We now extract the soft factor from the above expression as
\begin{align} \label{A_final}
&A(\{p_i\},\omega \hat{k}) =\sum_{i=1}^{n} \dfrac{\kappa}{2} \bigg(\dfrac{p_i^\alpha p_i^\beta \varepsilon_{\alpha\beta}}{\omega p_i\cdot \hat{k}} - i\dfrac{p_i^\alpha \varepsilon_{\alpha\beta} \hat{k}_\gamma J_i^{\beta\gamma}}{p_i\cdot \hat{k}} + \dfrac{1}{2\delta^2} \dfrac{p_i^\alpha p_i^\beta \varepsilon_{\alpha\beta}}{(p_i\cdot \hat{k})^2} + \dfrac{3}{2\delta^2} \dfrac{p_i^\alpha p_i^\beta \varepsilon_{\alpha\beta}}{(p_i\cdot \hat{k})^2} p_i\cdot\partial_{p_i} \nonumber \\
&- \dfrac{1}{\delta^2} \dfrac{(p_i^\alpha p_i^\beta \varepsilon_{\alpha\beta})(\Vec{p}_i\cdot \mathbf{\hat{k}})}{\omega (p_i\cdot \hat{k})^2} + \dfrac{1}{\delta^2} \dfrac{p_i^\alpha p_i^\beta \varepsilon_{\alpha\beta}}{p_i\cdot\hat{k}} \mathbf{\hat{k}}\cdot\partial_{\Vec{p}_i} - \dfrac{1}{\delta^2} \dfrac{(p_i^\alpha p_i^\beta \varepsilon_{\alpha\beta})(\vec{p}_i\cdot\mathbf{\hat{k}})}{(p_i\cdot\hat{k})^2} \hat{k}\cdot\partial_{p_i}
\bigg) \nonumber \\
&+ \dfrac{1}{\delta^2} \bar{N}^{\text{(dS)}}_{n+1}(\{p_1,...,p_n\}, \hat{k}),
\end{align}
where $N^{\text{(dS)}}_{n+1}=\bar{N}^{\text{(dS)}}_{n+1}\  \Gamma_n$. 

The first term inside the brackets is the Weinberg soft factor, the second term is the Cachazo-Strominger soft factor, and the remaining terms are the perturbative corrections.

\subsection{Results}
\label{results_sec}
The $\mathcal{S}$-matrix component $N^{\text{(dS)}}_{n+1}$ corresponding to the second diagram in Fig. (\ref{soft_graviton_fig}) where the graviton is not attached to an external line, does not receive $\frac{1}{\omega}$-pole terms \cite{SCB}. Thus, it does not introduce additional corrections to the Weinberg soft graviton theorem but to the Cachazo-Strominger soft theorem.

From Eq. (\ref{A_final}), we write the Weinberg soft graviton theorem including the perturbative corrections as follows:
\begin{equation} \label{soft_factor}
    \lim_{\omega\to 0}\ \omega A\big(\{p_i\},\omega\hat{k} \big) = \sum_{i=1}^{n} \dfrac{\kappa}{2}\ \dfrac{p_i^\alpha p_i^\beta \varepsilon_{\alpha\beta}}{p_i\cdot \hat{k}} \bigg(1-\dfrac{1}{\delta^2} \dfrac{\Vec{p}_i\cdot \mathbf{\hat{k}}}{p_i \cdot \hat{k}} \bigg).
\end{equation}
We can also write the Cachazo-Strominger soft theorem including the perturbative corrections as
\begin{align} \label{CS_soft_factor}
&\lim_{\omega\to 0}\ \big(1+\omega\partial_\omega \big)A(\{p_i\},\omega \hat{k}) = \sum_{i=1}^{n} \dfrac{\kappa}{2} \bigg(-i\dfrac{p_i^\alpha \varepsilon_{\alpha\beta} \hat{k}_\gamma J_i^{\beta\gamma}}{p_i\cdot \hat{k}} + \dfrac{1}{2\delta^2} \dfrac{p_i^\alpha p_i^\beta \varepsilon_{\alpha\beta}}{(p_i\cdot \hat{k})^2} + \dfrac{3}{2\delta^2} \dfrac{p_i^\alpha p_i^\beta \varepsilon_{\alpha\beta}}{(p_i\cdot \hat{k})^2} p_i\cdot\partial_{p_i} \nonumber \\ 
&+ \dfrac{1}{\delta^2} \dfrac{p_i^\alpha p_i^\beta \varepsilon_{\alpha\beta}}{p_i\cdot\hat{k}} \mathbf{\hat{k}}\cdot\partial_{\Vec{p}_i} - \dfrac{1}{\delta^2} \dfrac{(p_i^\alpha p_i^\beta \varepsilon_{\alpha\beta})(\vec{p}_i\cdot\mathbf{\hat{k}})}{(p_i\cdot\hat{k})^2} \hat{k}\cdot\partial_{p_i}
\bigg) + \dfrac{1}{\delta^2} \bar{N}^{\text{(dS)}}_{n+1}(\{p_1,...,p_n\}, \hat{k}).
\end{align}

To compute the perturbative corrections to the Cachazo-Strominger soft theorem, we need to determine $N^{\text{(dS)}}_{n+1}$ using the gauge invariance property explicitly in the momentum space. The gauge invariance at this order becomes highly complicated, which is out of the scope of the present work.

\section{Ward identity and perturbative soft graviton theorem}
\label{symmetry_sec}
In this section, we derive Ward identities corresponding to the perturbative soft graviton theorem (\ref{soft_factor}). We first express the soft factor in the holomorphic coordinates. The momenta and polarization tensor in these coordinates are parametrized as follows \cite{He_Lysov_Mitra_Strominger}:
\begin{equation}
    k^\mu = \omega\big(1,\mathbf{\hat{k}} \big),\ \ \mathbf{\hat{k}} = \bigg(\dfrac{w+\bar{w}}{1+w\bar{w}}, \dfrac{-i(w-\bar{w})}{1+w\bar{w}}, \dfrac{1-w\bar{w}}{1+w\bar{w}} \bigg)
\end{equation}
\begin{equation}
    p_i^\mu = E_i\big(1,\mathbf{\hat{p_i}} \big),\ \ \mathbf{\hat{p_i}} = \bigg(\dfrac{z_i+\bar{z_i}}{1+z_i\bar{z_i}}, \dfrac{-i(z_i-\bar{z_i})}{1+z_i\bar{z_i}}, \dfrac{1-z_i\bar{z_i}}{1+z_i\bar{z_i}} \bigg)
\end{equation}
\begin{equation}
    \varepsilon^{+\mu}(k) = \dfrac{1}{\sqrt{2}} \big(\bar{w},1,-i,-\bar{w} \big),\ \ \varepsilon^{-\mu}(k) = \dfrac{1}{\sqrt{2}} \big(w,1,i,-w \big).
\end{equation}
In this parametrization, the soft factor (\ref{soft_factor}) takes the form as
\begin{equation} \label{soft_factor_final}
    \lim_{\omega\to 0}\ \omega A\big(\{p_i\},\omega\hat{k} \big) = -\sum_{i=1}^{n} \dfrac{\kappa}{2}\ E_i \bigg[\bigg(1-\dfrac{1}{\delta^2} \bigg) \dfrac{(1+w\bar{w})(\bar{w}-\bar{z_i})}{(1+z_i\bar{z_i})(w-z_i)} + \dfrac{1}{2\delta^2} \dfrac{(1+w\bar{w})^2}{(w-z_i)^2} \bigg],
\end{equation}
which is written for the ``+"-helicity graviton, i.e. for $\varepsilon^+_{\mu\nu} = \varepsilon^+_{\mu} \varepsilon^+_{\nu}$.

\subsection{Supertranslation Ward identity: review}
\label{Ward_identity_review}
Given the asymptotic in- and out-states, $\ket{\text{in}}$ and $\ket{\text{out}}$, the Ward identity is defined by
\begin{equation}
    \braket{\text{out}|Q_f\mathcal{S}-\mathcal{S}Q_f|\text{in}} = 0,
\end{equation}
where $Q_f$ is the supertranslation charge which is expressed as
\begin{equation}
    Q_f = Q^{\text{soft}}_f + Q^{\text{hard}}_f,
\end{equation}
where
\begin{equation} \label{soft_charge_flat}
    Q^{\text{soft}}_f = \lim_{\omega\to 0} \dfrac{\omega}{2\pi} \int d^2z f(z,\bar{z}) D^2_{\bar{z}}a^{\text{flat}}_+(\omega,z,\bar{z}),
\end{equation}
\begin{equation} \label{hard_charge_flat}
    Q^{\text{hard}}_f = 2\int du d^2z \gamma_{z\bar{z}} f(z,\bar{z}) T_{uu}.
\end{equation}
Therefore, the supertranslation Ward identity can be written as follows \cite{Campiglia_Laddha_2015}:
\begin{equation}
    \lim_{\omega\to 0} \dfrac{\omega}{2\pi} \int d^2z f(z,\bar{z}) D^2_{\bar{z}} \braket{\text{out}|a^{\text{flat}}_+(\omega,z,\bar{z})\mathcal{S}|\text{in}} = -\sum_{i=1}^{n} E_i f(z_i,\bar{z_i}) \braket{\text{out}|\mathcal{S}|\text{in}},
\end{equation}
where $a^{\text{flat}}_+$ is the annihilation operator\footnote{The operators defined on the de Sitter background reduce to those of flat space operators in the limit of $\delta\to\infty$, i.e. $\lim\limits_{\delta\to\infty}a_{\pm}(\omega,z,\bar{z}) = a^{\text{flat}}_\pm(\omega,z,\bar{z})$.} that acts on the asymptotic states at null infinity in the Minkowski space.

One can recover the Weinberg soft graviton theorem by taking a particular choice of $f$ in the Ward identity. To do so, we first perform the integration by parts on the LHS of the above expression:
\begin{equation} \label{Ward_flat_2}
    \lim_{\omega\to 0} \dfrac{\omega}{2\pi} \int d^2z D^2_{\bar{z}} f(z,\bar{z}) \braket{\text{out}|a^{\text{flat}}_+(\omega,z,\bar{z})\mathcal{S}|\text{in}} = -\sum_{i=1}^{n} E_i f(z_i,\bar{z_i}) \braket{\text{out}|\mathcal{S}|\text{in}}.
\end{equation}
Let us now consider the choice of $f$ as \cite{Campiglia_Laddha_2015}
\begin{equation} \label{f_choice}
    f(z,\bar{z}) = s(z,\bar{z};w,\bar{w}) = \dfrac{(1+w\bar{w})(\bar{w}-\bar{z})}{(1+z\bar{z})(w-z)},
\end{equation}
where $D^2_{\bar{z}}s(z,\bar{z};w,\bar{w})=2\pi\delta^{(2)}(z-w)$.
Similarly, on the RHS of Eq. (\ref{Ward_flat_2})
\begin{equation} \label{f_choice_zi}
    f(z_i,\bar{z_i}) = s(z_i,\bar{z_i}; w,\bar{w}) = \dfrac{(1+w\bar{w})(\bar{w}-\bar{z_i})}{(1+z_i\bar{z_i})(w-z_i)}.
\end{equation}
Thus, the Ward identity (\ref{Ward_flat_2}) now reduces to
\begin{equation}
    \lim_{\omega\to 0} \omega \braket{\text{out}|a^{\text{flat}}_+(\omega,w,\bar{w})\mathcal{S}|\text{in}} = -\sum_{i=1}^{n} E_i \dfrac{(1+w\bar{w})(\bar{w}-\bar{z_i})}{(1+z_i\bar{z_i})(w-z_i)} \braket{\text{out}|\mathcal{S}|\text{in}},
\end{equation}
which is the Weinberg soft graviton theorem\footnote{The constant factor $\kappa/2$ was omitted.} expressed in the holomorphic coordinates.

\subsection{Perturbative corrections to supertranslation Ward identity}
\label{perturbative_Ward_sec}
In this section, we derive a supertranslation Ward identity corresponding to the perturbative soft graviton theorem. We first put the perturbative soft graviton theorem (\ref{soft_factor_final}) in the bracket notations as
\begin{multline} \label{perturbative_soft_bracket}
    \lim_{\omega\to 0} \omega \braket{\text{out}|a_+(\omega,w,\bar{w})\mathcal{S}|\text{in}} \\
    = -\sum_{i=1}^{n} E_i \bigg[\bigg(1-\dfrac{1}{\delta^2} \bigg) \dfrac{(1+w\bar{w})(\bar{w}-\bar{z_i})}{(1+z_i\bar{z_i})(w-z_i)} + \dfrac{1}{2\delta^2} \dfrac{(1+w\bar{w})^2}{(w-z_i)^2} \bigg] \braket{\text{out}|\mathcal{S}|\text{in}},
\end{multline}
where the constant factor $\kappa/2$ has been omitted.

Let us now apply the operator $(2\pi)^{-1}\int d^2 w f(w,\bar{w})D^2_{\bar{w}}$ on both sides of the above expression.
\begin{multline} \label{Ward_perturbative_1}
    \lim_{\omega\to 0} \dfrac{\omega}{2\pi} \int d^2w f(w,\bar{w}) D^2_{\bar{w}}\braket{\text{out}|a_+(\omega,w,\bar{w})\mathcal{S}|\text{in}} \\
    = -\sum_{i=1}^{n} \dfrac{E_i}{2\pi} \int d^2w f(w,\bar{w}) D^2_{\bar{w}} \bigg[\bigg(1-\dfrac{1}{\delta^2} \bigg) \dfrac{(1+w\bar{w})(\bar{w}-\bar{z_i})}{(1+z_i\bar{z_i})(w-z_i)} + \dfrac{1}{2\delta^2} \dfrac{(1+w\bar{w})^2}{(w-z_i)^2} \bigg] \braket{\text{out}|\mathcal{S}|\text{in}}
\end{multline}
Before applying the covariant derivative $D_{\bar{w}}$ on the RHS of the above expression, we must note the tensor structure of the annihilation operator $a_+$ on the LHS. The tensor structure of $a_+$ can be read from the expression $a_+ = \frac{2\pi i}{\sqrt{\gamma}}C_{zz}$ given in Eq. (9) of \cite{Campiglia_Laddha_2015}, where $C_{zz}$ is a radiative data at null infinity. The covariant derivative on $a_+$ acts as follows:
\begin{align}
    &D_{\bar{w}} a_+(\omega,w,\bar{w}) = \gamma^{w\bar{w}} \partial_{\bar{w}}\big(\gamma_{w\bar{w}}a_+(\omega,w,\bar{w}) \big), \\
    &D^2_{\bar{w}} a_+(\omega,w,\bar{w}) = \partial_{\bar{w}} \big(\gamma^{w\bar{w}} \partial_{\bar{w}}(\gamma_{w\bar{w}}a_+(\omega,w,\bar{w})) \big).
\end{align}
Thus, the covariant derivative on the soft factor on the RHS of Eq. (\ref{Ward_perturbative_1}) acts in a similar way as described above. Therefore,
\begin{align}
    & D^2_{\bar{w}}\bigg[\dfrac{(1+w\bar{w})(\bar{w}-\bar{z_i})}{(1+z_i\bar{z_i})(w-z_i)} \bigg] = 2\pi\delta^{(2)}(w-z_i), \label{Dw2_identity} \\
    & D_{\bar{w}}\bigg[\dfrac{(1+w\bar{w})^2}{(w-z_i)^2} \bigg] = -4\pi \gamma^{w\bar{w}}\partial_w\delta^{(2)}(w-z_i) \label{Dw_on_correction},
\end{align}
where the identity $\partial_{\bar{w}}\dfrac{1}{(w-z_i)^2} =-2\pi\partial_w\delta^{(2)}(w-z_i)$ has been used to obtain (\ref{Dw_on_correction}).
Now, we use the identities (\ref{Dw2_identity}) and (\ref{Dw_on_correction}) to evaluate the integrals on the RHS in Eq. (\ref{Ward_perturbative_1}). The first integral is trivial due to the presence of the delta function. To evaluate the second integral, we first perform the integration by parts in $w$ and then use the identity (\ref{Dw_on_correction}):
\begin{align}
    \dfrac{1}{2\pi}\int d^2w f(w,\bar{w}) D^2_{\bar{w}} \bigg[\dfrac{(1+w\bar{w})^2}{(w-z_i)^2} \bigg] 
    &= -\dfrac{1}{2\pi}\int d^2w D_{\bar{w}} f(w,\bar{w}) D_{\bar{w}} \bigg[\dfrac{(1+w\bar{w})^2}{(w-z_i)^2} \bigg] \nonumber \\
    &= 2\int d^2w D_{\bar{w}} f(w,\bar{w}) \gamma^{w\bar{w}}\partial_w\delta^{(2)}(w-z_i) \nonumber \\
    &=-2\partial_{z_i} D^{z_i} f(z_i,\bar{z_i}).
\end{align}
Thus, Eq. (\ref{Ward_perturbative_1}) reduces to
\begin{multline} \label{Ward_perturbative_2}
    \lim_{\omega\to 0} \dfrac{\omega}{2\pi} \int d^2w f(w,\bar{w}) D^2_{\bar{w}}\braket{\text{out}|a_+(\omega,w,\bar{w})\mathcal{S}|\text{in}} \\
    = -\sum_{i=1}^{n} E_i \bigg[\bigg(1-\dfrac{1}{\delta^2} \bigg) f(z_i,\bar{z_i}) - \dfrac{1}{\delta^2} \partial_{z_i} D^{z_i}f(z_i,\bar{z_i}) \bigg] \braket{\text{out}|\mathcal{S}|\text{in}}.
\end{multline}
We rewrite the above expression in the $(z,\bar{z})$ coordinates as
\begin{multline} \label{Ward_perturbative_final}
    \lim_{\omega\to 0} \dfrac{\omega}{2\pi} \int d^2z f(z,\bar{z}) D^2_{\bar{z}}\braket{\text{out}|a_+(\omega,z,\bar{z})\mathcal{S}|\text{in}} \\
    = -\sum_{i=1}^{n} E_i \bigg[\bigg(1-\dfrac{1}{\delta^2} \bigg) f(z_i,\bar{z_i}) - \dfrac{1}{\delta^2} \partial_{z_i} D^{z_i}f(z_i,\bar{z_i}) \bigg] \braket{\text{out}|\mathcal{S}|\text{in}},
\end{multline}
which is the supertranslation Ward identity corresponding to the perturbative soft graviton theorem (\ref{perturbative_soft_bracket}). The corresponding supertranslation charges are
\begin{equation} \label{soft_charge_dS}
    Q^{\text{soft}}_f = \lim_{\omega\to 0} \dfrac{\omega}{2\pi} \int d^2z f(z,\bar{z}) D^2_{\bar{z}}a_+(\omega,z,\bar{z}),
\end{equation}
\begin{equation} \label{hard_charge_dS}
    Q^{\text{hard}}_f = 2\int du d^2z \gamma_{z\bar{z}} f(z,\bar{z}) T_{uu} - \dfrac{2}{\delta^2} \int dud^2z \gamma_{z\bar{z}} \big[f(z,\bar{z}) + \partial_z D^z f(z,\bar{z}) \big] T_{uu},
\end{equation}
which are consistent with the flat space limit $\delta\to\infty$.

\subsection{Perturbative soft graviton theorem from Ward identity}
\label{Ward_to_soft_sec}
In this section, we show and verify that a particular choice of $f$ in the supertranslation Ward identity derived in the previous section gives the perturbative soft graviton theorem. We first perform integration by parts on the LHS of Eq. (\ref{Ward_perturbative_final}).
\begin{multline}
    \lim_{\omega\to 0} \dfrac{\omega}{2\pi} \int d^2z D^2_{\bar{z}} f(z,\bar{z}) \braket{\text{out}|a_+(\omega,z,\bar{z})\mathcal{S}|\text{in}} \\
    = -\sum_{i=1}^{n} E_i \bigg[\bigg(1-\dfrac{1}{\delta^2} \bigg) f(z_i,\bar{z_i}) - \dfrac{1}{\delta^2} \partial_{z_i} D^{z_i}f(z_i,\bar{z_i}) \bigg] \braket{\text{out}|\mathcal{S}|\text{in}}.
\end{multline}
We choose the same $f$ as in (\ref{f_choice}). Thus, the covariant derivatives of $f$ on the LHS give the delta function, and its integration reduces to the LHS of the perturbative soft graviton theorem (\ref{perturbative_soft_bracket}). On the RHS of the above expression, the first term matches with that of (\ref{perturbative_soft_bracket}) which can be seen from Eq. (\ref{f_choice_zi}). And the second term reduces to
\begin{equation}
    \partial_{z_i} D^{z_i} f(z_i,\bar{z_i}) = -\dfrac{(1+w\bar{w})^2}{2(w-z_i)^2},
\end{equation}
which also matches with that of (\ref{perturbative_soft_bracket}). Thus, we have verified that the Ward identity (\ref{Ward_perturbative_final}) for the same choice of $f$ as in (\ref{f_choice}) reduces to the perturbative soft graviton theorem (\ref{perturbative_soft_bracket}).

\section{Discussion}
\label{discussion_sec}
In this paper, we studied the tree-level scattering of massless scalars with an emission of a soft graviton in the small compact region inside the static patch of the de Sitter space, and derived the perturbative soft graviton theorem and the corresponding supertranslation Ward identity. We first obtained the orthogonal set of mode solutions for the massless scalar field in the limit of large curvature length $l$. Using this orthogonal set of modes, we defined the scalar field propagator, and verified that it is indeed the Green's function of the massless scalar field equation. We then defined and evaluated the $\mathcal{S}$-matrix for the soft graviton scattering process shown in Fig. (\ref{soft_graviton_fig}), and derived the perturbative corrections to the Weinberg soft graviton theorem. Further, we derived the supertranslation Ward identity corresponding to the perturbative soft graviton theorem, and showed that the perturbative soft graviton theorem can be recovered from the supertranslation Ward identity for the same choice of supertranslation parameter $f$ as in flat space.

The supertranslation charges with perturbative corrections, i.e. Eqs. (\ref{soft_charge_dS}) and (\ref{hard_charge_dS}), have been deduced from the Ward identity (\ref{Ward_perturbative_final}) derived in this paper. These charge expressions reduce to those of the usual supertranslation charges, i.e. Eqs. (\ref{soft_charge_flat}) and (\ref{hard_charge_flat}), in the flat space limit $\delta\to\infty$.

The perturbative corrections to the Ward identity and the supertranslation charges in this paper have been derived from the perturbative soft graviton theorem. Thus, a natural question arises whether the same corrected charges can be derived from an asymptotic symmetry analysis. The asymptotic symmetry analysis requires a null infinity structure, which is absent in the de Sitter space. This obstacle can be circumvented in the limit of large curvature length of de Sitter space as follows: We consider the de Sitter metric in static coordinates, which in the limit of large curvature length can be treated as the background Minkowski space with a small perturbation in the Bondi gauge as $h_{\mu\nu}=\frac{r^2}{l^2}\delta^{u}_{\mu}\delta^{u}_{\nu}$. This perturbation vanishes at small distances in the interior, and the spacetime reduces to the Minkowski space. At large distances, the perturbation remains small and finite due to a double scaling limit in which the ratio $\frac{r}{l}\ (<<1)$ is held fixed as $r\to\infty$ and $l\to\infty$, which is consistent with the soft limit discussed in section (\ref{setup_sec}). Thus, it can be expected that the symmetry generators and supertranslation charges at null infinity in Minkowski space receive corrections due to this non-trivial perturbation. Performing the asymptotic symmetry analysis with appropriate boundary conditions to derive corrected supertranslation charges is indeed very interesting problem and we leave it for future work.

Moreover, as discussed in section (\ref{results_sec}), finding the perturbative corrections to the Cachazo-Strominger soft theorem requires one to determine the subleading correction term $\bar{N}^{\text{(dS)}}_{n+1}$ in Eq. (\ref{CS_soft_factor}). This term is extremely difficult to obtain using the current method due to the complicated structure of the gauge transformations at subleading order $\mathcal{O}(l^{-2})$. An alternative approach is to determine the perturbative corrections to the flat space superrotation or sphere vector field Ward identity first, and then deduce the corrected Cachazo-Strominger soft factor from it. Extending the asymptotic symmetry analysis at subleading order and incorporating the double scaling limit may cause several challenges, which are currently unclear to us.

\section*{Acknowledgements}
D.N.S. is extremely grateful to Alok Laddha for many insightful discussions on connecting the perturbative soft graviton theorem to the Ward identity. D.N.S. also gratefully acknowledges the hospitality of the Chennai Mathematical Institute where part of this work was done. D.N.S. further thanks Srijit Bhattacharjee for carefully reviewing the draft and for his valuable suggestions.

\end{document}